\documentclass [winedt,yap]{iitparc}
\usepackage{cite}
\usepackage{amsmath,amssymb,amsfonts,bm}
\usepackage{lscape}
\usepackage{floatfig,wrapfig,epsfig}
\usepackage{subfigure}
\usepackage{color}
\usepackage{psboxit}
\usepackage{rotating}
\usepackage{curves}
\usepackage{ulem}
\usepackage[mathscr]{eucal}
\RequirePackage{psfrag}
\RequirePackage{graphicx}

\begin{document}\normalem
\initfloatingfigs
\frontmatter          

\IssuePrice{25.00}%
\TransYearOfIssue{2010}%
\TransCopyrightYear{2010}%
\OrigYearOfIssue{2010}%
\OrigCopyrightYear{2010}%

\TransVolumeNo{71}%
\TransIssueNo{6}%
\OrigIssueNo{20}%


\mainmatter

\setcounter{page}{1196}
\CRubrika{CONTROL IN SOCIAL ECONOMIC SYSTEMS} \Rubrika{CONTROL
IN SOCIAL ECONOMIC SYSTEMS}
%
\OrigJournalName{Problemy Upravleniya}
\OrigIssueNo{4}
\OrigYearOfIssue{2008}
\OrigCopyrightYear{2008}%

\title{Analysis of Collectivism and Egoism Phenomena\\ within the Context of Social Welfare}
\titlerunning{Analysis of Collectivism and Egoism Phenomena}

\author{P. Yu.~Chebotarev, A. K.~Loginov, Ya. Yu.~Tsodikova, \\ Z. M.~Lezina, and V. I.~Borzenko}
\authorrunning{Chebotarev  \lowercase{et\ al.}}
\OrigCopyrightedAuthors{P.Yu.~Chebotarev, A.K.~Loginov,
Ya.Yu.~Tsodikova, Z.M.~Lezina, V.I.~Borzenko}

\institute{Trapeznikov Institute of Control Sciences, Russian
Academy of Sciences, Moscow, Russia}
\received{Received March 11, 2008}
\OrigPages{pp.~30--37}

\maketitle

\begin{abstract}
Comparative benefits provided by basic social strategies, i.e.,
collectivism and egoism, are investigated within the framework
of democratic collective decisions.
\end{abstract}

\DOI{0202}

\section{INTRODUCTION TO A PROBLEM}

Egoism of a voter means supporting the proposals that meet his
individual interests. On the other hand, collectivism is to
protect the interests of a certain group associated with the
voter in question. The egoist's problem lies in the fact that,
having a single vote, he is often unable to protect personal
interests. A group has more chances to succeed through solidary
voting; however, a member of the group still faces a problem.
In particular, the larger (more powerful) is the group, the
greater can be the differences between the interests of the
group and the individual ones of its members. Therefore, the
question ``What is more beneficial for a person, i.e., playing
for himself or protecting interests of the group that may not
coincide with individual ones?'' is nontrivial and has no
universal answer. Similar problems arise for political parties
and movements; they should choose possible candidates for
making unions and alliances, as well as evaluate opportunities
for autonomous participation in the political process.

In the case of a small group uniting participants with similar
interests, collectivism of the members gets close to egoism.
Contrariwise, in a large group that approximates the size of a
society, collectivism turns out to be a version of altruism; in
fact, every participant protects the interests common for the
society.

Problems of relating egoism, altruism and rationality have been
studied in [1--3]; voting as a method to make decisions
regarding distribution of social benefits via income taxation
and implementation of social programs has been considered in
[4--7]. Democratic decisions made by egoistic voters are easily
manipulated by means of formulation of proposals; this fact has
been demonstrated in the theory of voting. In this context we
should mention investigations performed, first, by
A.V.~Malishevskii (1970) of the Institute of Control Sciences
[8, pp.~92--95] and, second, by R.~McKelvey (1976) of the
California Institute of Technology [9]; in addition, the reader
is referred to [10].

For instance, the result of a sequence of decisions made by
\textit{overwhelming majority} may be disadvantageous for
\textit{all voters without exception}. Such manipulation is
impossible with respect to collectivists seeking to maximize
the benefits of large groups. Hence, the following type of
social dynamics is an attractive area of research. Some
participants form a group in order to protect their common
interests. Assume they succeed in solving this task; i.e., the
average group member gets more benefits than an egoist. Since
the group is open for new members, the group size increases. If
the group remains successful under the growth, then there is no
obstacle to further expansion, up to the size of the community;
during the growth, the egoism of the group approaches altruism.
As Arthur Koestler states, ``The egoism of the group feeds on
the altruism of its members;'' some years earlier, the same
idea was expressed by Reinhold Niebuhr. The phenomenon under
study is an instance of the inverse transformation.

We have analyzed the corresponding effects using mathematical
models in several papers. In particular, it has been shown in
[11--13] that, for large domains in the parameter space of the
model under consideration, the group can successfully (in terms
of the utility of its members) compete with individual
participants. This fact points to the practicability of the
growing ``snowball'' of cooperation, the type of social
dynamics mentioned above. At the same time, under certain
conditions, the group looses, and it is more profitable to
behave individually.

In order to further investigate the benefits and effective
mechanisms of cooperative behavior, one should consider the
case where the group competes not only with individual
participants, but also with other groups (``parties''). The
subject of the present paper is to analyze the case of two
rival groups.

\section{MODEL AND STATEMENT OF THE PROBLEM}

A model suitable for the study of the collectivist and egoistic
strategies should possess the following features. It should be
sufficiently rich to describe the basic elements of reality
being modeled and sufficiently simple to allow the use of
analytical methods.

The framework of the model should allow to formalize the
following key notions: the interests of participants,
cooperation, the interests of groups, profitability, variable
environment, social opportunities and challenges, making and
implementing collective decisions.

The basic assumptions of the model we propose are as follows.
\begin{itemize}
  \item []
\begin{enumerate}
\item[(1)] The current state of every participant is
    characterized by a quantitative indicator that may
    be viewed as the level of welfare, capital,
    utility, social role, satisfaction, success, etc.
    For brevity, we use a capital-like interpretation
    of this indicator in the sequel. The indicator may
    have both positive and negative values. The model
    allows to specify the initial values of the
    indicator for all the participants.
\item[(2)] An egoistic participant is interested in
    maximizing his capital.
\item[(3)] Participants may form groups.
\item[(4)] The success of a group is described by
    nondecreasing functions in the capitals of the
    members.
\item[(5)] The social dynamics under study is
    determined by collective decisions made through
    voting involving all participants.
\item[(6)] To make decisions, the $\alpha$-majority
    procedure is used; the parameter $\alpha$
    determines the number of votes necessary and
    sufficient to accept a proposal. E.g.,
    ${\alpha}=0.5$ describes the procedure of simple
    majority vote. If a proposal is rejected,
    \textit{status quo\/} is preserved, i.e., the
    capitals of the participants remain the same.
\item[(7)] While voting, every egoistic participant
    supports precisely the proposals increasing his
    capital (this item makes item~(2) more exact).
\item[(8)] The members of each group vote together. We
    consider several group criteria of supporting
    proposals, \textit{viz.}, (a)~the criterion of
    $\alpha $-majority (with a threshold ${\alpha}_1 $
    that may differ from the general vote threshold)
    and (b) a sufficiently high level of the average
    capital increment for the members of the group
    (e.g., exceeding~0).
\item[(9)] Proposal is modeled by the vector of capital
    increments of all the participants.
\item[(10)] As a generator of proposals we consider
    realizations of a random vector with independent
    identically distributed components. The proposals
    are interpreted as opportunities produced by a
    stochastic environment, which can be favorable,
    neutral or unfavorable.
\item[(11)] As the distribution of a capital increment
    we use Gaussian distribution $N({\mu ,}\,{\sigma
    }^2)$. Parameters $\mu $ and $\sigma $ characterize
    the environment, i.e., ${\mu }>0$, ${\mu }=0$ and
    ${\mu }<0$ correspond to a favorable, neutral and
    unfavorable environment, respectively; $\sigma$
    stands for the variability of the environment.
\item[(12)] Various modifications of the model are
    analyzed. They differ in the conditions of joining
    and leaving groups, as well as in the conditions of
    leaving the game for those participants who went
    bankrupt (have negative capital accounts), possible
    influence of the capital value on the capital
    increment, etc. In addition, other voting rules and
    principles to generate and support proposals are
    considered.
\end{enumerate}
\end{itemize}

The main subject of investigation is social dynamics defined by collective decisions under the assumptions listed above. In particular, we analyze
\begin{itemize}
\item[$\cbd$] comparative benefits of different social
    strategies including egoism and collectivism with
    respect to both the participants and the whole society;
\item[$\cbd$] relations between the results and the
    parameters of the model, including the levels of favor
    and variation in the environment, voting thresholds,
    supporting principles for the proposals, the ratio of
    participants with different strategies, and so on.
\end{itemize}

Special attention is paid to the interpretation of the results
in social terms.

This paper is mostly dedicated to the case where participants
are divided into three categories, namely, egoists and the
members of two groups (``parties'').

\section{``SNOWBALL'' OF COOPERATION}

One of the major problems of social reality can be formulated
as follows. Egoistic behavior of social agents often leads to
rather regrettable results. Falling into trouble or having a
need of support, a pure egoist gets no assistance. In addition,
faced with the slightest bit organized opposition, the egoist
usually turns out to be powerless. Generally speaking,
altruistic behavior is more profitable for the society until it
is typical for the absolute majority of participants.\
Otherwise, altruists only have a sense of moral right while the
remaining dividends are distributed among less noble members.
In other words, altruism provides a beneficial equilibrium
which is extremely unstable. In modern society, one can observe
altruism that in the least could be considered wide-spread, in
the sphere of symbolic actions; they include demonstration of
courtesy, attentiveness (sometimes in the way reminiscent of
the popular ignorant heroes of Russian literature, Mr.\
Bobchinskii and Mr.\ Dobchinskii, described by the famous
Russian writer N.V.~Gogol'). Other examples of altruism can be
found in areas where the interests of the participant in
question are not appreciably affected (for instance,
philanthropy). With coming commercialization, even the ``noble
professions'' gradually lose their special status. We should
mention important exceptions such as the rules of conduct in
emergency (in a shipwreck, women and children should be saved
first, etc.), although not always these rules are followed. In
general, altruism remains the prerogative of noble people,
whose number is few in any era.

It appears that no changes towards altruism can be found in
international politics. As before, foreign policy of any
government is to protect the interests of the state it
represents; this often leads to cynical attitude towards the
interests of other nations. Domestic policy has similar
tendencies when representatives of elected authorities, even
protecting the interests of the electorate, are often willing
to ignore interests of other citizens.

The above political examples give samples of corporate
(cooperative) behavior. A~representative of a state (or a
community, or a corporation) protects the interests of his
community that competes with the others. We should emphasize
that consideration of the complete range of social strategies
from egoism to altruism makes it possible to give examples of
cooperative behavior that correspond to every segment of this
range.

As has been noted, the case of cooperative behavior within an
opened and competitive (and, thus, increasing) union of
participants, deserves particular attention. As soon as the
union increases, cooperative behavior tends to altruistic one.
This social mechanism may be referred to as ``\textit{snezhnyi
kom kooperatsii}'' (``snowball'' of cooperation). In jest,
setting hopes on it may be termed ``\textit{snezhnyi
kom}''-munizm (``snowball communism''). In the following
sections, we study the model described in Section~2.

\section{WHEN IS THE ``SNOWBALL'' GROWING AND WHEN IS IT MELTING?}

\vspace*{1mm}
First of all, consider a society consisting of social agents of
two types, egoists and group members. Let us ask a question on
how the benefits of the group, the egoists and the whole
society depend on group size.

Formulas obtained in [12] provide a complete answer to this
question. The approximate expressions presented in the theorem
below are less cumbersome than the precise ones; the
approximation introduces acceptably small errors, with the only
exception in the case of small groups, where we use accurate
formulas. In the following theorem, a pair of values in square
brackets stands for a row matrix or a column matrix, while
$\lfloor\theta n\rfloor$ denotes the integer part of $\theta
n$.

\medskip
\textbf{Theorem~}[12]. {\it Let $M(\tilde {d}_{\varepsilon } )$
and $M(\tilde {d}_G)$ be the mathematical expectations of the
one-step capital increments for the egoists and group
members$,$ respectively. Suppose that the group supports a
proposal if and only if this proposal ensures a positive
increment of the total group capital. Then the standard normal
approximation of the binomial distribution provides the
following expressions$:$
\begin{gather}
\begin{split}
& M(\tilde {d}_{\varepsilon })\approx [P_G\;\,Q_G]\,\left(
{\mu\left[{\begin{array}{l}
 F_\gamma\\
 F_\alpha    \\
 \end{array}}\right]+\frac{\sigma f}{\sqrt{pq\ell}}\left[
{\begin{array}{l}
 f_\gamma\\
 f_\alpha\\
 \end{array}} \right]} \right),\\[.6em]
& M(\tilde {d}_G )\approx[F_\gamma\;\,F_\alpha]\,\left(
{\mu\left[{\begin{array}{l}
 P_G\\
 Q_G\\
 \end{array}} \right]+\frac{\sigma f_G}{\sqrt{g}}\left[
{\begin{array}{l}
 \;\;{\kern 1pt}{\kern 1pt}1\\
 -1 \\
 \end{array}}\right]}\right),
\end{split}
\end{gather}

\noindent where
${P_G=F\left({\dfrac{\mu\sqrt{g}}{\sigma}}\right)}$,
${ Q_G=1-P_G =F\left({-\dfrac{\mu\sqrt{g}}{\sigma}} \right)}$,
${F_\theta=F\left({-\dfrac{\lfloor\theta n\rfloor+0.5-p\ell
}{\sqrt{pq\ell}}}\right)}$,
${f_\theta=}\linebreak {f\left({\dfrac{\lfloor\theta n\rfloor+0.5-p\ell
}{\sqrt{pq\ell}}} \right)}$,
${\theta\in\left\{\alpha,\gamma\right\}}$,
${\gamma=\alpha-g/n}$,
${p=F\left(\mu/\sigma\right)}$,
${q=1-p}$,
${f_G =f(\mu\sqrt{g}/\sigma)}$,\linebreak
${f=f\left(\mu/\sigma\right)}$,
$F(\cdot)$ and $f(\cdot)$ are the standard normal distribution
function and the corresponding density function$,$ $\ell$ and
$g$ denote the number of egoists and the group size$,$
respectively$,$ and $n=\ell +g$ is the number of participants}.

\medskip
A typical example of the relationship between one-step capital
increments of the participants and the group size is given by
Fig.\,1. In this example, the society consists of 1000 agents;
the $x$-axis refers to the group size (in the range between 0
and 1000). The group members vote as stated in the theorem
above. The parameters of the distribution of proposals are:
${\mu }=-0{.}8$ and ${\sigma }=30$; the voting threshold is
${\alpha }=0{.}5$. On the ordinate, the average (expected)
one-step capital increments for a group member, an egoist and a
randomly selected participant are plotted.

\begin{figure}[t]
\vspace*{1mm}

\centerline{\includegraphics[scale=1.05]{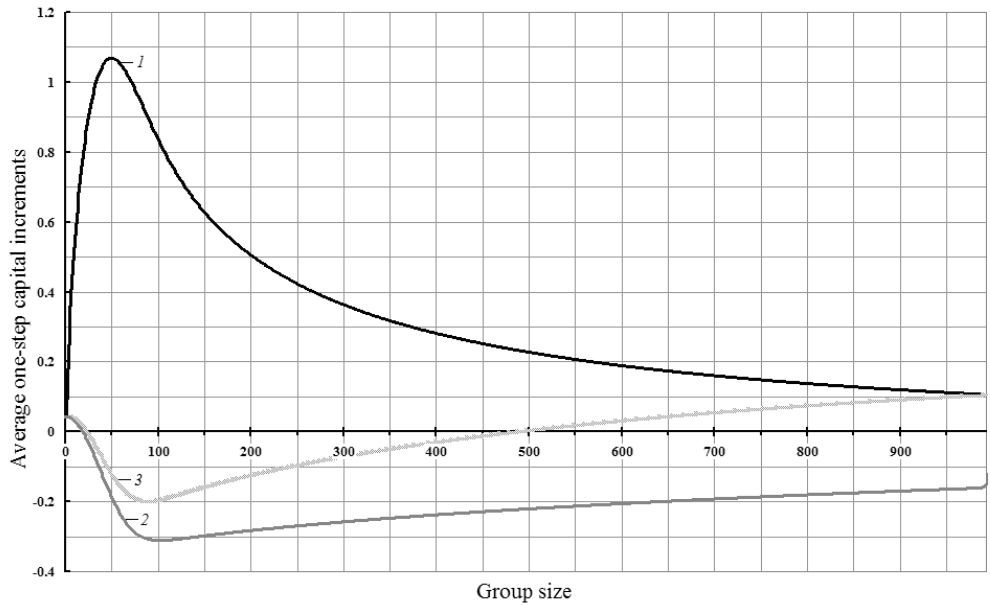}}
\label{fig1}

\vspace*{1mm}
\caption{Average one-step capital increments for: ({\sl1\/}) a group member, ({\sl2\/}) an egoist
and ({\sl3\/}) a randomly selected participant: 1000 participants (the group and egoists); $\mu=-0{.}8$,
$\sigma =30$, $\alpha =0{.}5.$}
\end{figure}

It should be noted that the group, irrespective of its size,
has a significant advantage over the egoists. Therefore, it is
profitable for the egoists to join the group. The members of a
small group gain a lot of benefits from increasing the group
size. On the contrary, expansion of a small group is not
profitable for the egoists and the whole society. In the
example under consideration, the most profitable (for the group
members) size of the group is 50 persons. Further expansion of
the group leads to the decrease of the average capital
increments of the participants. However, starting from 88
members, further expansion of the group becomes profitable for
the whole society, and starting from 102 members, it is
beneficial for the egoists either. The whole society enters the
domain of positive average capital increments with 488 members
of the group and it reaches the one-step capital increment
typical of the group-free case when the group has 649 members.

This leads to the following conclusion, which is relevant in
the design of control algorithms. If the ``rules of play'' are
established by those interested in maximizing the total capital
of the society, then the group should be open to new members.
Then a likely scenario is that all participants join the group
and thereby the group egoism becomes a version of altruism.
Note that this altruism is conditional: should any member start
voting for his personal interests conflicting with the
interests of the group, and the rest of the group will stop
protecting this member's interests. In other words, ``all for
one'' as long as ``one for all.'' It is precisely this
condition that ensures the \emph{stability\/} of
``cooperative'' (reciprocal, in the terminology of Robert
Trivers) altruism; this stability distinguishes it from the
``absolute'' one (also called hard-core altruism).

Now imagine that the ``rules of play'' are determined by the
participants themselves. In this case, every participant is
interested in such a minimum ceiling of group size that exceeds
49 and guarantees the presence of this participant within the
group. If the group already has 50 members, then they will seek
to draw a line (i.e., do not invite new members). In this case,
the group constitutes a ``solid elite'' that resembles the
party in a one-party political system. The existence of such a
group is unprofitable to the whole society; but the worse is
when the group exceeds its (internally) optimal size. In the
example under consideration, the minimum value of the total
social capital is reached when the group contains 88 members.

Certainly, in addition to attempts to join the group, which is
\textit{de facto} an ``elite,'' there exists an alternative and
more natural strategy:\ it consists in organizing a competitive
group. The consequences are discussed in Section~5.

\section{THE SECOND ``SNOWBALL''}

Let the group reach its internally optimal size; assume that
further expansion is blocked. Consider the organization of the
second group. The analysis techniques we use involve, on the
one hand, a simulation model implemented on a computer and, on
the other hand, analytical expressions analogous to Eqs.\,(1)
and too cumbersome to be presented here. Figure~2 demonstrates
how the expected one-step capital increments of the egoists,
the members of two groups and randomly chosen participants
depend on the size of the second group.

\begin{figure}[b]
\centerline{\includegraphics[scale=1.05]{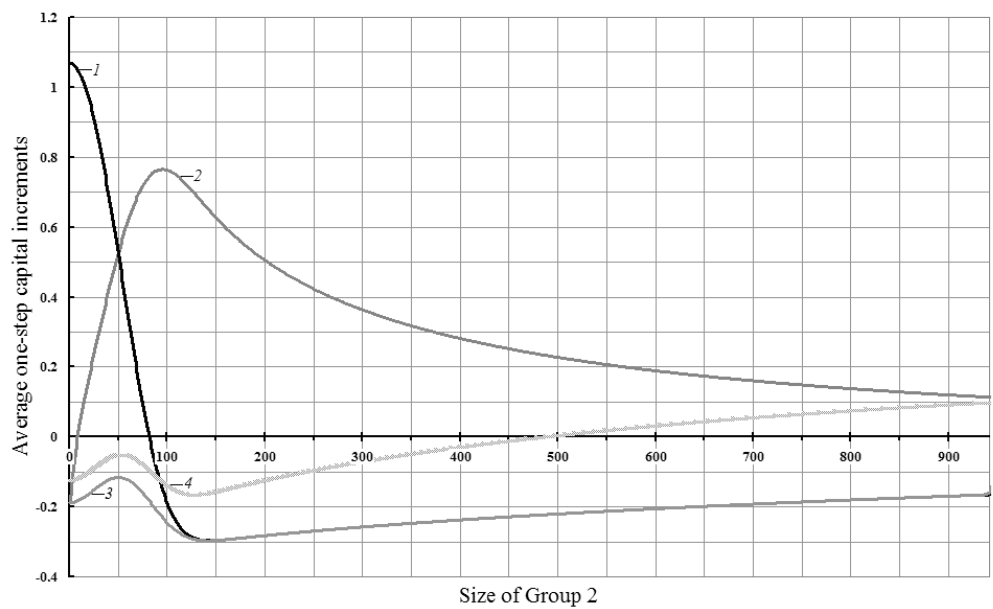}}
\label{fig2}
\caption{Average one-step capital increments for: ({\sl1})~a member of the first group,
({\sl2})~a member of the second group, ({\sl3})~an egoist, ({\sl4})~a randomly selected participant.
The community of 1000 participants includes 50~members in the first group, the second group and egoists; $\mu =-0{.}8$,
$\sigma =30$, $\alpha =0{.}5.$}
\end{figure}

It is easily seen that the expansion of the second group is
extremely disadvantageous for the first one: it causes a swift
decrease of the average capital increments of the first group.
The capital increment of the second group member, \textit{per
contra\/}, grows reaching its maximum level at 96 members. In
the present example, this maximum level is 72\% of that for the
sole first group. If the second group further expands via the
addition of egoists, the advantage of the first group over the
egoists vanishes; this happens as the size of the second group
approaches 150 members. The continued addition of the egoists
to the second group (which reduces its ``elitism'') is
profitable for them, for the remaining egoists, for the first
group and for the whole society. The first group attains its
minimum level of average capital increment when the second one
has 144 members.

Thus, for a single group, it is beneficial to have a small
size; if two groups compete, it is the larger one that wins.
However, the winning group is not interested in exceeding the
other one by a factor of more than 2--2{.}5.

Let us stress the following point. In the presence of two
competing groups, the egoists benefit the most when the groups
compete ``as equals;'' i.e., the average capital increment of
an egoist is maximal when the groups have the same size. A
political counterpart of this situation is the two-party system
with the parties having the same vote share. In the above
example, such a situation is optimal for the ``unorganized''
part of the society. The mechanism of this phenomenon is as
follows: the closer are the parties in their power, the more
they value the votes of ``unorganized'' citizens. Accordingly,
the greater is the impact of those votes on the decisions being
made. In the present example, the situation where one of the
groups considerably outnumbers the other is even slightly worse
for the egoists than the presence of only one group having the
combined size of those two.

\section{TWO ``SNOWBALLS'' THAT GROW TOGETHER}

In Section 5 we have established the fact that an existing
group loses to another one that has just appeared and has a
higher rate of size increase. A~conclusion is: with the
appearance of the second group, the first one should not
``freeze'' its size. On the contrary, it should strive to
maintain its advantage in size. Consider the case where two
groups are characterized by the same rate of size increase,
with the first group having 50 extra members. A corresponding
example is shown in Fig.\,3.

\begin{figure}[b]
\centerline{\includegraphics[scale=1.05]{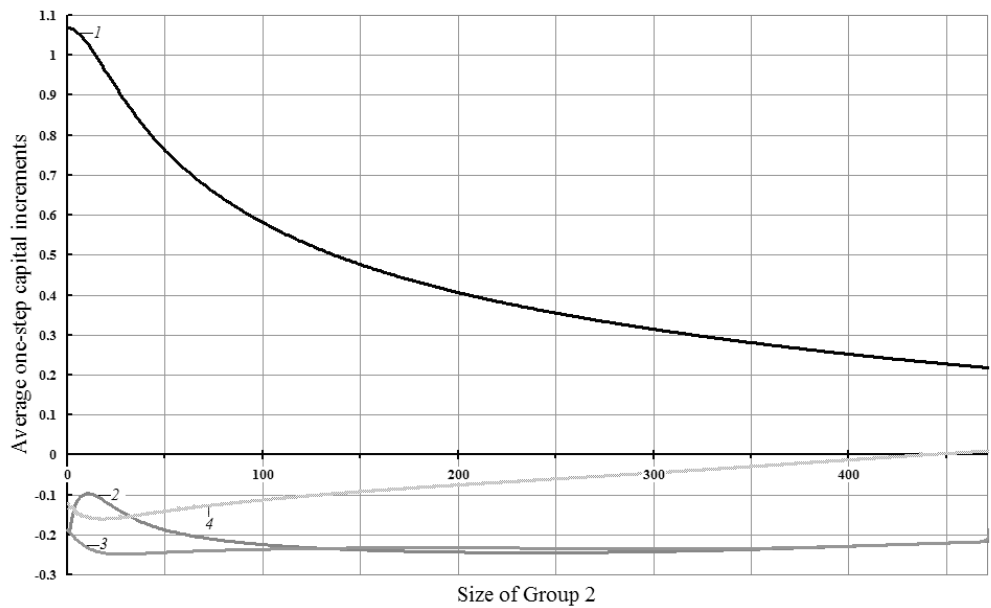}}
\caption{Average one-step capital increments for: ({\sl1}) a member of the first group, ({\sl2}) a member of the second group, ({\sl3}) an egoist and ({\sl4}) a randomly selected participant. The society consists of 1000 participants; the first group has 50 extra participants compared to the second one; $\mu =-0.8,\;\sigma =30,\;\alpha =0.5.$}
\label{fig3}
\end{figure}

As could be expected, the first group succeeds in keeping a
dominant position over the second one using such a policy. With
the increase of the group size, the average capital increment
of a Group~1 member is reduced five times, but it still remains
much higher than the increments of the other participants
(which do not leave the domain of negative values). During this
process, the increment of the whole society goes up, because
the leading group becomes less ``elite.'' In the very
beginning, the average capital increment of a Group~2 member
grows swiftly; however, it soon reaches its maximum
(approximately, $-0.1$ at 10 members) and starts to decrease,
reaching the minimum value of $-0.245$. Within this segment,
the benefit of the second group is even less than that of the
egoists; with further growth of the groups, it slightly
increases. In fact, the second group goes broke since it loses
to the more powerful first group in votes almost always. The
difference in 50 votes is so significant that only in extremely
rare cases it can be compensated by the votes of the egoists.

In contrast to this, if the difference between the groups is
small (say, 5 members instead of 50), then the dynamics is
quite different, cf.\ Fig.\,4.

\begin{figure}[t]
\centerline{\includegraphics[scale=1.05]{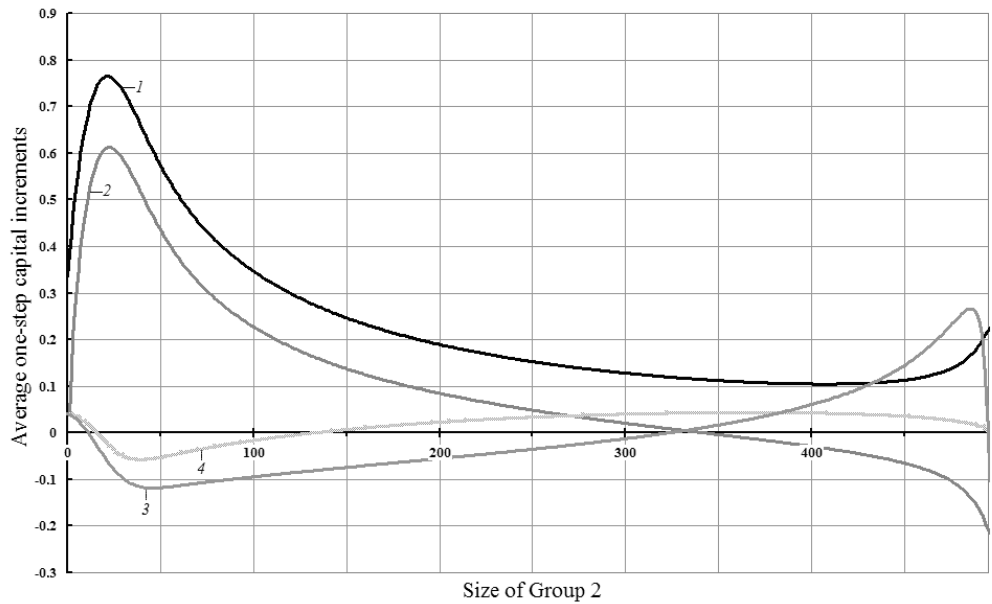}}
\caption{Average one-step capital increments for: ({\sl1}) a member of the first group, ({\sl2}) a member of the second group, ({\sl3}) an egoist and ({\sl4}) a randomly selected participant. The society consists of 1000 participants; the first group has 5 extra participants compared to the second one; $\mu =-0.8,\;\sigma =30,\;\alpha =0.5.$}
\label{fig4}
\end{figure}

In addition to other status of the second group, we would like
to attract the reader's attention to the curve of the average
capital increment of an egoist: in particular, it has a
``rise'' and a ``drop'' in the right-hand side of the diagram
in Fig.\,4. A~well-known effect consisting in the advantages of
a ``small party'' in the presence of two greater ones having an
approximate balance can be observed here. The egoists are
uncoordinated; nevertheless, when they are few and none of the
groups is able to secure a majority, the society mainly accepts
the proposals supported by a substantial proportion of the
egoists. Naturally, this situation is profitable for them. This
explains the ``rise'' of the average capital increment of the
egoists and their advantage over both groups in the
corresponding segment. However, when the number of egoists is
less than 5, all decisions are simply made by the first group;
thus the capital increment of this group goes up while the
curves describing the benefits of the other participants go
down. Interestingly, the minimum capital increment of a society
member is reached when the second group has the size (i.e., 40
participants) approximately twice that (though the absolute
difference is moderate) providing the maximum capital
increments for the groups (22 and 23 members, respectively).
This dynamics is similar to that shown in Fig.~1. The maximum
level of the society's capital increment is observed when the
two groups, as well as the egoists, compete most intensively
and ``as equals'' (the corresponding size of the second group
is 367 members). When the size of the second group reaches 384,
the benefit of the egoist ``catches up'' the one of a randomly
selected participant. The formulas we derived allow one to
elucidate the mechanism of the above phenomena (as well as some
others). However, keeping within the limits of this paper makes
it impossible to discuss the point in detail.

\section{TWO ``BALLS OF WOOL''}

Now let us consider the case where redistribution of
participants between two groups does not involve egoists, i.e.,
the sum of the two group sizes, as well as the number of
egoists, remain constant. A corresponding example is shown in
Fig.\,5; the $x$-axis refers to the size of the first group;
with the size of the second group it adds up to 1000. The
society consists of the two groups and 500 egoists. Other
parameters are: $\mu=-0.8$, $\sigma=100$, $\alpha=2/3$
(a~qualified majority).

In contrast to ``snowball,'' this scenario resembles the
hourglass or two balls of wool with a shared thread,
\textit{viz.}, what is wound off the first ball is wound on the
second one.

\begin{figure}[b]
\centerline{\includegraphics[scale=1.05]{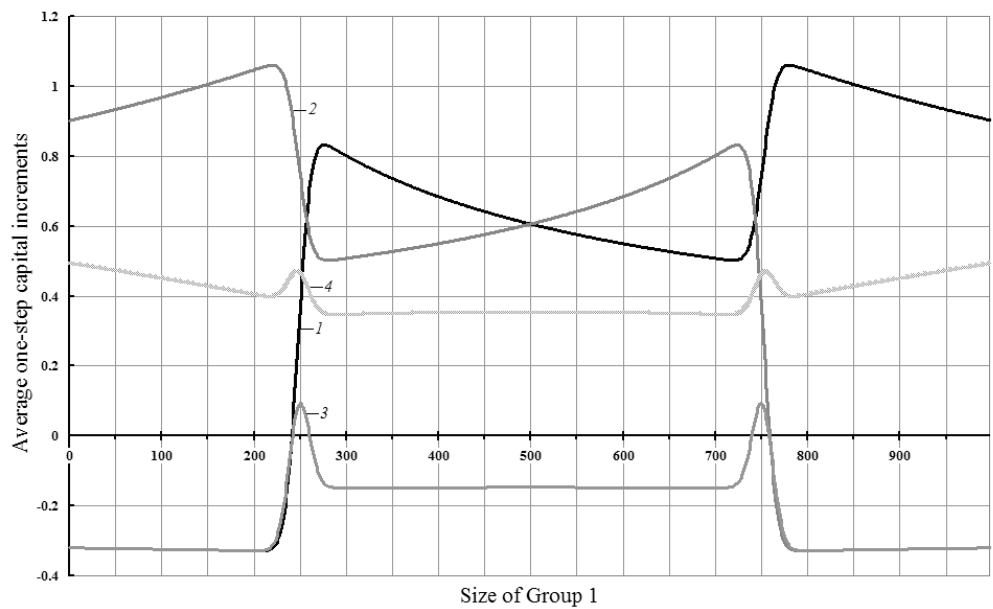}}
\label{fig5}
\caption{Average one-step capital increments for: ({\sl1}) member of the first group, ({\sl2}) member of the second group, ({\sl3}) egoist and ({\sl4}) randomly selected participant: 1500 participants, the first and the second group have 1000 members totally; $\mu =-0.8,\;\sigma =100,\;\alpha =2/3.$}
\end{figure}

The results presented in Fig.\,5 are somewhat unexpected; below we comment them.

To accept a proposal, 1001 votes are needed. If the first group
has more than 780 members and supports a proposal, then the
remaining votes (no more than 220) are usually provided by the
egoists; the proposal cannot be accepted without the support of
the first group. Therefore, in this size segment, the first
group has a high level of average capital increments. If the
size of the first group is increased in the range of 780 to
1000 members, then its average capital increment is reduced.
The reason is that the larger is the group, the smaller the
frequency of its satisfaction with the proposals will be. Along
with this, the average capital increments of its members
measuring this satisfaction are decreased. If the group
includes 720 to 780 members, then, provided that it supports a
proposal, the egoists often fail to supply the remaining votes.
However, if the egoists manage to bring those votes, then the
proposal, which is quite profitable for them, will be accepted.
That is why the segment in question is remarkable for top
capital increments of the egoists. Moreover, when the size of
the first group belongs to [720,~750] and this group supports a
proposal, the lacking votes are often provided by the second
group, which gains from this; therefore the capital increments
of the second group are significant here. In the segment
[750,~780], the remaining votes are more and more frequently
provided by the egoists, which results in a great drop of the
capital increment of the second group. In [280,~720], almost
all accepted proposals are supported by both groups. This
situation is more favorable for the smaller one since,
according to the law of averages, its satisfaction, in the
mean, is expressed by greater levels of capital increment. At
the same time, this segment is more favorable for the egoists
than [780,~1000] as far as here, the number of accepted
proposals having negative average capital increments for the
egoists is smaller. The cause is that here, to accept such a
proposal, the agreement of both groups is required, whereas for
the segment [780,~1000], the support of the first group is
sufficient. The segment [220,~280] is symmetric to [720,~780].
The segment [0,~220] is characterized by the fact that almost
all proposals favorable to the second group are accepted here
and so the first group and the egoists cannot protect their
interests.

Let us note that the two-party system with the parties having
equal powers is very stable in the present example. Indeed, if
two groups have 450 and 550 members, then the members of the
larger group will gain some benefits from joining the smaller
one since the latter has advantages. This leads to the size of
the groups becoming almost equal with the lapse of time.
However, the stable two-party system is not the most profitable
for the egoists and the whole society. The matter is that, due
to a high voting threshold, the votes of the egoists do not
allow any party to secure the acceptance of a beneficial
proposal without the support by the second party. Therefore,
the interest of egoists is actually ignored here. Their votes
are highly demanded when the groups have approximately 750 and
250 members. In this case, to accept a proposal beneficial to
the egoists, it is sufficient to enlist the support of the
larger group. On the contrary, to accept a proposal
disadvantageous (at the average) to the egoists, both groups
must support it, which is far less probable. Thus the egoists
have the maximum capital increments when the first group
includes 250 or 750 members.

\section{CONCLUSION}

The real state of things is much more intricate than the simple
model considered in this paper; in particular, there exist many
subjective factors that have not been taken into account within
the framework of the model. Nevertheless, analysis of the model
allows identification of certain hidden mechanisms of social
processes. The model can serve as a ``zero-order
approximation'' that highlights many relevant phenomena, e.g.,
benefits of the two-party system with parties having almost the
same power, advantages of a small party in the presence of two
rival large parties, and so on. The model may be implemented
with respect to a society, a parliament, a collective, etc.,
provided that the distribution of power and the decision-making
procedures are known; analysis of the model enables one to
understand the special features of the social system under
study and, possibly, to optimize its performance. The results
of such analysis are sometimes difficult to forecast; even a
few examples discussed above give an indication of the numerous
types of system performance that can be implemented by varying
the model parameters.

The ``snowball'' of cooperation is the central metaphor of the
paper; it refers to the group that efficiently protects the
interests of its members and is open to new members. Through
its expansion, such a group becomes more altruistic.

It turns out, however, that, starting from a certain group
size, further expansion of the group may become disadvantageous
for its members by causing a decrease of the capital
increments. From this point on, the interests of the group and
of the remaining participants diverge. This phenomenon was
studied in Section~4; it is related to many actual economic and
political situations (for example, to the problem of dividing
the Arctic shelf among the competing states).

When the members of a group are not interested in its
expansion, a rival group can be organized. As was demonstrated
in Section~5, the primary group loses if during this process it
tries to ``freeze'' its optimal size. Contrariwise, to maintain
a dominant position, this group should grow; the rate of this
growth should be moderate (see Section~6). The case of two
large groups having almost the same size turns out to be
beneficial for the ``unorganized'' members of the society,
provided that the decisions are made by simple majority. For
the whole society, the most beneficial situation is reaching
consensus; in the absence of consensus the society should aim
at the situations of real competition involving all
participants. This conclusion should be taken into account
while designing control algorithms for social and economic
systems.

Finally, Section~7 was devoted to the analysis of the
``interchange'' of members between two groups. It might seem
surprising that the members of a smaller group have more
benefits than the members of the larger one. Such a phenomenon
can ensure the stability of the two-party system with parties
of almost the same size. The ``unorganized'' part of the
society naturally has the highest benefit when its votes are
highly demanded. It has been shown that such a situation occurs
not only in the two-party system with equal parties; some
unbalanced situations may also be profitable for the egoists,
especially, under qualified majority voting systems.

This paper does not claim to provide a systematic description
of all types of social dynamics implemented within the
framework of the model; the limits of a short article make such
a description impossible. Our task was to consider a few
typical examples which are relevant to social issues.

Getting back to the ``snowball'' of cooperation, let us emphasize
that there exist many cases where the members of an alliance have
no reason to ``freeze'' its size. In particular, this is apparently
typical of international cooperation on global problems such as the problem of reducing emissions into the atmosphere. 
The members of such an alliance implement joint projects for
the sake of the alliance and, to some extent, of the other
counties. Since this activity does not contradict the interests
of the third parties, the alliance is typically interested in
its further expansion. On the other hand, in situations where
an alliance (an ``elite'') is concerned with limiting its size,
its openness can be established as a standard as far as it
meets the interests of the society. In all cases mentioned
above the mechanism of ``snowball'' of cooperation can be used
to create stable ``conditionally altruistic'' communities (see
Section~4). Notably, this mechanism can be involved in the
development of civil society, which is one of the main problems
of today (see, e.g., [14]). This relies on the fact that the
cells of civil society benefit from joining associations that
maintain a high level of solidarity and coordinate their
actions.

\revred{F.T. Aleskerov}

\end{document}